# Dynamics of Charge Flow in the Channel of a Thin-Film Field-Effect Transistor

E.G. Bittle, J.W. Brill, and J.P. Straley

Department of Physics and Astronomy

University of Kentucky, Lexington, KY 40506-0055, USA

Abstract

The local conductivity in the channel of a thin-film field-effect transistor is proportional to the charge density induced by the local gate voltage. We show how this determines the frequency- and position-dependence of the charge induced in the channel for the case of "zero applied current": zero drain-source voltage with charge induced by a square-wave voltage applied to the gate, assuming constant mobility and negligible contact impedances. An approximate expression for the frequency dependence of the induced charge in the center of the channel can be conveniently used to determine the charge mobility. Fits of electro-optic measurements of the induced charge in organic transistors are used as examples.

## I. Introduction

Reflecting the growing interest in developing devices utilizing organic semiconductors and in understanding their properties, probes are being developed that can measure the spatial dependence of semiconductor properties within the channel of thin-film field effect transistors, in which all the mobile charge is induced by applied voltages. Scanning Kelvin probe microscopes are being used to measure the variation of the potential[1-3] and optical probes are being used to measure the distribution and dynamics of charge in the channel.[4-10] In particular, dynamic measurements[6,7,10] are going beyond the usual, "lumped-impedance" approximation of a field-effect transistor (FET). We've recently reported[7] measurements of the frequency dependence of spatially resolved infrared reflectance in an organic, small molecule field-effect transistor when a square-wave voltage was applied to the gate. In this paper, we give a theory that describes how an FET charges and discharges and compare it with our measurements.

## II. Model and Analysis

For our model, we assume a bottom-contact geometry in which the semiconductor has a constant width between the drain and source electrodes (with channel length L), that all the semiconductor charge is at the interface with the dielectric, and that there is negligible leakage current through the gate dielectric.[5] In addition, we neglect fringing fields, so that the problem is one-dimensional, with the local potential $V(x,t)$ simply proportional to the local charge density, $V(x,t) = \rho(x,t) / C$, where C and $\rho$ are the capacitance (to the gate) and charge per unit area. (This model is therefore inappropriate for top-contact FETs,[11-13] or even bottom contact transistors for which a barrier adjacent to the contact affects charge injection.[14]) Assuming that current density in the channel between the drain and source is limited by diffusion and conductance in the semiconductor,

$$j = - D\, \partial\rho/\partial x - \mu\rho\, \partial V/\partial x = - (D + \mu\rho/C)\, \partial\rho/\partial x , \quad (1)$$

$$\partial\rho/\partial t = - \partial j/\partial x = (D + \mu\rho/C)\, \partial^2\rho/\partial x^2 + \mu/C\, (\partial\rho/\partial x)^2, \quad (2)$$

where D is the carrier diffusion constant and $\mu$ the mobility.[10] Lin *et al*[13] have solved Eqtn. (2) to find the drain current in the case $D \approx 0$ (see below) with field dependent $\mu$ and with contact impedances as appropriate for a top-contact FET. Instead, we take $\mu$ as constant, as discussed below. As boundary conditions, we assume that there are negligible contact impedances and injection barriers, so that at the source and drain, $\rho(0,t) = CV_{gs}(t)$ and $\rho(L,t) = CV_{gd}(t)$, as is often appropriate for bottom contact transistors.[14] Before turning to the charge dynamics, we point out that for the static case with $D \approx 0$, $V_{gd} = 0$, and $V_{gs} \neq 0$, the charge density has the form $\rho(x) = CV_{gs}\,(1-x/L)^{1/2}$, as observed in Reference [8].

We now turn to the dynamics when the drain-source voltage is kept at 0 V, so that there is no applied current and all current is induced by a changing gate voltage. For square-wave voltages applied to the gate[6,7] (i.e. $V_{gs}$ switching between 0 and $V_{gs0} \equiv \rho_0/C$), Eq. (2) can be written as

$$\partial\rho/\partial t = (D + \mu\rho_0/2C)\, \partial^2\rho/\partial x^2 + (\mu/C)\, [(\rho - \rho_0/2)\, \partial^2\rho/\partial x^2 + (\partial\rho/\partial x)^2]. \quad (3)$$

This equation can be solved numerically for different frequencies and values of $D/\mu V_{gs0}$. Note, however, that since one expects $D \sim \mu k_B T/e$, the diffusion term would be negligible near room temperature for typical applied voltages $V_{gs0} > 1$ V. We therefore only present our solutions in the $D \approx 0$ limit. The time dependence of $\rho$ at a few square-wave frequencies, normalized to $F_0 \equiv \mu V_{gs0}/2L^2$, and at a few positions (shown in the inset to Figure 2) are shown in Figure 1. Also shown, for comparison, is the linear response, i.e. neglecting the term in brackets in Eq. (3) (the heat flow equation), at the center to the channel, $x = L/2$.

One striking aspect of the non-linear responses is that the FET discharges more slowly than it charges. This is because when the gate voltage turns off, the charge density adjacent to the contacts immediately becomes zero, so the conductivity ($\mu\rho$) adjacent to the contact also becomes zero, and there is a bottleneck for charge to leave the interior of the channel. (This bottleneck effect is not significantly changed by including a very small but finite D, e.g. $D = 0.02 \; \mu V_{gs0}$.) This asymmetry between charging and discharging was noticed and discussed by Matsui and Hasegawa.[6] A consequence is that the average charge becomes greater than $\rho_0/2$ (the value for linear response).

While the discharge of the transistor is slower than one would expect from linear response, the charging speed is greater, so that the average time constants for linear and non-linear response are expected to be similar, as discussed further below. Also note that the time dependence of the responses at $x = L/2$ and $x = L/3$ are very similar, especially at lower frequencies. (At higher frequencies, the response at $x = L/2$ is slightly smaller and delayed from that at $x = L/3$.) This similarity is a result of the fact that charge is flowing from contacts on both ends of the sample. A useful consequence is that if one uses measurements of the charge near the center of the channel to analyze the charge dynamics, as done below, it is not essential that measurements be done precisely in the center.

For small signals, it is usually easier to measure the frequency dependence of signals proportional to the charge modulation (e.g. with a lock-in amplifier) than the time dependence.[6,7] We have therefore calculated the fundamental component ($\rho_F$) of the Fourier expansion of $\rho(t)$ at the three locations for several values of square-wave frequency, as shown in Figure 2. The responses both in phase and in quadrature with the applied square-waves are shown. Also shown are these values calculated for the linear response at $x=L/2$. We note the following:

a) The non-linear responses are very similar at $x = L/2$ and $x = L/3$, as observed[6] and discussed above. Qualitatively, they can be characterized as over-damped harmonic oscillators; in particular, at high frequencies the in-phase response becomes inverted,[6,7] as charge is still flowing into (out of) the center when the voltage is turned off (on), as is also seen in the $F = 16 \; F_0$ graphs of Figure 1.

b) On the other hand, at $x = L/6$, there is a long tail extending to high frequencies, reflecting the rapid diffusion of charge from the nearby contact.

c) The quadrature peaks of the non-linear and linear responses at $x = L/2$ occur at almost the same frequencies. The biggest difference between the non-linear and linear responses is that the quadrature peak (and the step in the in-phase response) are broader for the non-linear case, indicating that if one was to fit these curves as overdamped oscillators, one would need a larger distribution of relaxation times for the non-linear response.[15] The larger width for the non-linear response reflects the fact that the charging and discharging times differ.

While fits of the response at the center of the channel to an overdamped oscillator with a distribution of time constants are possible,[15] it is more convenient to use a simpler, rational expression which can be used to fit experimental data to determine the mobility. Figure 2 (bold-faced curves) show fits of both the in-phase and quadrature responses at the center of the channel to Eq. 4; the parameters for the fits are given in Table I.

$$\rho_F(L/2)/\rho_0 \approx (\alpha - \gamma z + \varepsilon z^2) / (1 - \beta z + \delta z^2), \text{ with } z \equiv \ln(F/F_0) = \ln(2FL^2/\mu V_{gs0}). \quad (4)$$

We emphasize that Equation (4) is an approximate expression, but one that can conveniently be used to fit charge oscillations in the center of the channel.

### III. Experimental Fits and Discussion

A fit of Eq. 4 to the frequency dependence of our infrared electro-optic data of Reference [7] (with $F_0$ and the overall scale as fitting parameters) is shown in Figure 3a. In this experiment, we measured the change in reflectance at the center of the channel of a 6,13 bis(triisopropylsilylethynyl)-pentacene (TIPS-Pn)[16] bottom contact FET as a function of the frequency of a square-wave voltage applied to the gate. The dielectric was the cross-linked polymer poly(vinylphenol) (PVP),[17] ~ 1 μm thick, spin-cast on a gold gate electrode. The channel length was L ≈ 400 μm with five TIPS-Pn crystals spanning the gold source and drain electrodes; complete electro-optic measurements were only done on one crystal. A negative gate voltage pulled holes onto the surface of the TIPS-Pn crystals, increasing their absorption and therefore decreasing the signal reflected from the gate, with the change in reflectance $|\Delta R/R|$ proportional to $|V_{gs0}|$.[7] Although the data is noisy, a good fit to Eq. (4) is obtained with $F_0$ = 630 Hz at $V_{gs0}$ = -50 V, giving a mobility μ ≈ 0.04 cm$^2$/V·s. This is in reasonable agreement with the linear mobility, $\mu_{lin}$ ≈ 0.06 cm$^2$/V·s, found from the IV curve of the FET, because the latter depends on the capacitance, which was not precisely known, and is also an average value for all five crystals spanning the electrodes. In fact, mobilities varying by a factor of two have been observed for crystals on similar transistors, as shown below. As emphasized in Reference [7], the electro-optic measurement provides an estimate of the mobility without knowledge of the capacitance, effective area, or total charge.

In fact, we estimated a charge coverage of only ≈ 0.4% for the FET of Figure 3a with $V_{gs}$ = -50 V .[7] Unfortunately, we cannot improve the signal/noise ratio of our electro-optic data by reducing the thickness of the dielectric (and increasing ρ) because of the "buried metal layer" effect.[18] That is, for reflectance from the metallic gate, the maximum absorption of light of wavelength λ occurs for a dielectric film of thickness $t_0$ ≈ λ/4n, where n = the index of refraction; for PVP, n ≈ 2,[16] so for λ ≈ 11 μm, $t_0$ ≈ 1 μm. However, we found that for some FETs, there were small regions (i.e. smaller than our spatial resolution, 10 μm, in diameter) where we observed very large electro-optic signals. In fact, ΔR was positive for some such "hot-spots" (e.g. see Figure 3b), suggesting that the electro-optic response was determined by an increase in reflectance rather than an increase of absorption of the charged layer; perhaps a grain of TIPS-Pn is in a pin-hole in the PVP and very close to the gate. ($|\Delta R/R|$ varied as $V_{gs0}^2$ at the hot-spot, suggesting that the capacitance of the grain also increased with gate voltage.)

The frequency dependence of the response at one such "hot-spot" located near the center of a crystal of length L ≈ 375 µm, with $V_{gs0}$ = -50 V, is shown in Figure 3b, together with a fit to Eq. (4). The excellent fit indicates that, whatever the mechanism for the enhanced electro-optic response at the "hot-spot," the FET is charging/discharging as "normal", i.e. charge is flowing along the crystal surface with constant mobility and capacitance as described by Eq. (3). Indeed, we fit the electro-reflectance at a position on the crystal close to but off the hot-spot with the same value of $F_0$ = 2300 Hz, as shown in Figure 3c, corresponding to a mobility µ ≈ 0.13 $cm^2$/ V·s. The mobility calculated from the IV curve for this transistor, $µ_{lin}$ ≈ 0.06 $cm^2$/ V·s, was somewhat smaller, suggesting that some of the other crystals spanning the electrodes did not carry much current, e.g. due to cracks or poor contacts. Indeed, we were able to measure the frequency dependence of the electro-reflectance of another crystal in the same FET, also shown in Figure 3c, and it had a much smaller value of $F_0$ (700 Hz) and mobility (0.07 $cm^2$/ V·s). (Note that this crystal crossed the channel on an angle and was therefore longer, L = 500 µm.)

Figure 3c points out one of the advantages of measurements of the frequency dependence of the local charge density or potential (by any technique): fits to Eqtn. (4) can be used for *in situ* determination of the mobility of different crystals in the same transistor. Deviations from the fit can be associated with the breakdown of an assumption such as constant mobility or negligible contact impedance.

Our assumption of constant µ, i.e. independent of j and ρ, deserves more comment. Meyertholen *et al*[5] have shown that the mobility increased rapidly with charge density in a polymer FET, and associated this dependence with disorder-induced shallow traps. This would increase the "bottleneck" effect for discharging FETs as well as affect the spatial dependence of charge. Small molecule, crystalline semiconductors are certainly expected to have less disorder than semiconducting polymers, and such has been demonstrated for rubrene.[19] Different results have been obtained for pentacene transistors; references [10,13] report strongly field dependent mobilities for FETs with silica dielectrics while the spatial dependence of the charge reported by Manaka *et al*[8] for pentacene on a polymer dielectric is consistent with constant mobility. Our fits to the frequency dependence of TIPS-Pn transistors suggest that disorder induced traps are also not important in these devices, as also shown in Reference [20]. In addition, our fits show that, reflecting the long channels and resulting long drift times of our transistors, contact impedances do not significantly affect their dynamic response.

In conclusion, we have presented a one-dimensional nonlinear equation describing the spatial/temporal variation of charge in an FET channel as it charges and discharges. The results, assuming constant mobility, are in qualitative agreement with measurements of the charge and potential variation in channels[1,6-8] and in quantitative agreement with our measurements for the frequency dependence of charge in the center of a channel when a square-wave voltage is applied to the gate.[7] For the latter, we have provided an approximate rational expression that can be used to determine the mobility.

We thank John Anthony for providing TIPS-Pn and helpful discussions. This work was supported in part by the U.S. National Science Foundation through Grants DMR-0800367 and EPS-0814194 and the Center for Advanced Materials.

|   | **in-phase** | **quadrature** |
|---|---|---|
| $\alpha$ | 0.3561 | 0.2541 |
| $\beta$ | 0.3174 | 0.5293 |
| $\gamma$ | 0.3375 | 0.03513 |
| $\delta$ | 0.1448 | 0.3202 |
| $\varepsilon$ | 0.06726 | -0.01693 |

**Table 1.** Parameters of Equation (4) for the in-phase and quadrature response.

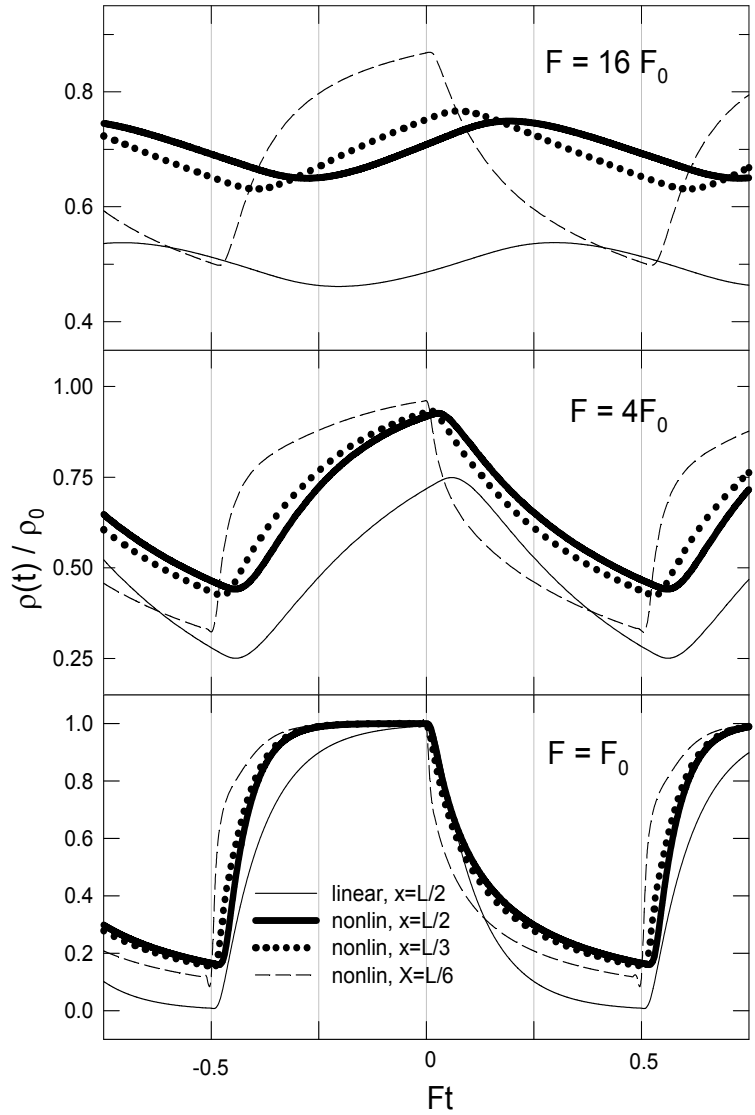

**Figure 1.** Time dependence of the charge density for an applied square-wave at $x = L/6$, $L/3$, and $L/2$ (center of channel) for three frequencies, determined by solution of Eqtn. (3) with diffusion constant $D = 0$. L is the length of the crystal and x is the distance from the source, as shown in the inset to Figure 2, $F_0 \equiv \mu V_{gs0}/2L^2$, where $V_{gs0}$ = the square-wave amplitude, and $\rho_0$ is the charge density at the contacts ($x = 0, L$) during the "on" half-cycle. Also shown is the linear response at $x = L/2$.

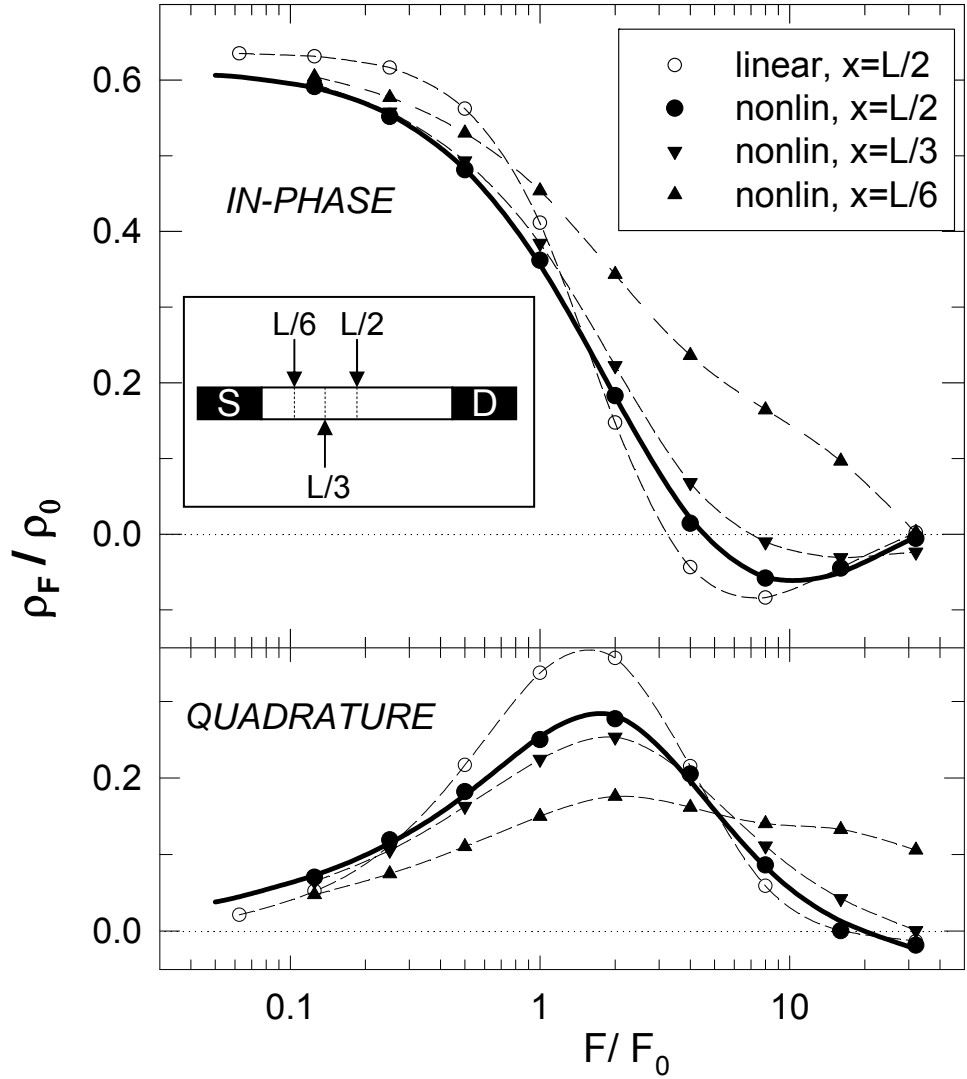

**Figure 2.** The frequency dependence of the fundamental Fourier component of the square-wave response (solution of Eqtn. (3) with D=0) at x = L/6, L/3, and L/2. Also shown is the transform of the linear response at x = L/2. The responses in phase (upper panel) and in quadrature (lower panel) with the applied square-wave are shown. For the non-linear response at x = L/2, the bold-faced curve shows the fit given by Eqtn. 4; for the other sets, the curves are guides to the eye only. The inset shows the three locations on the channel where the response is calculated.

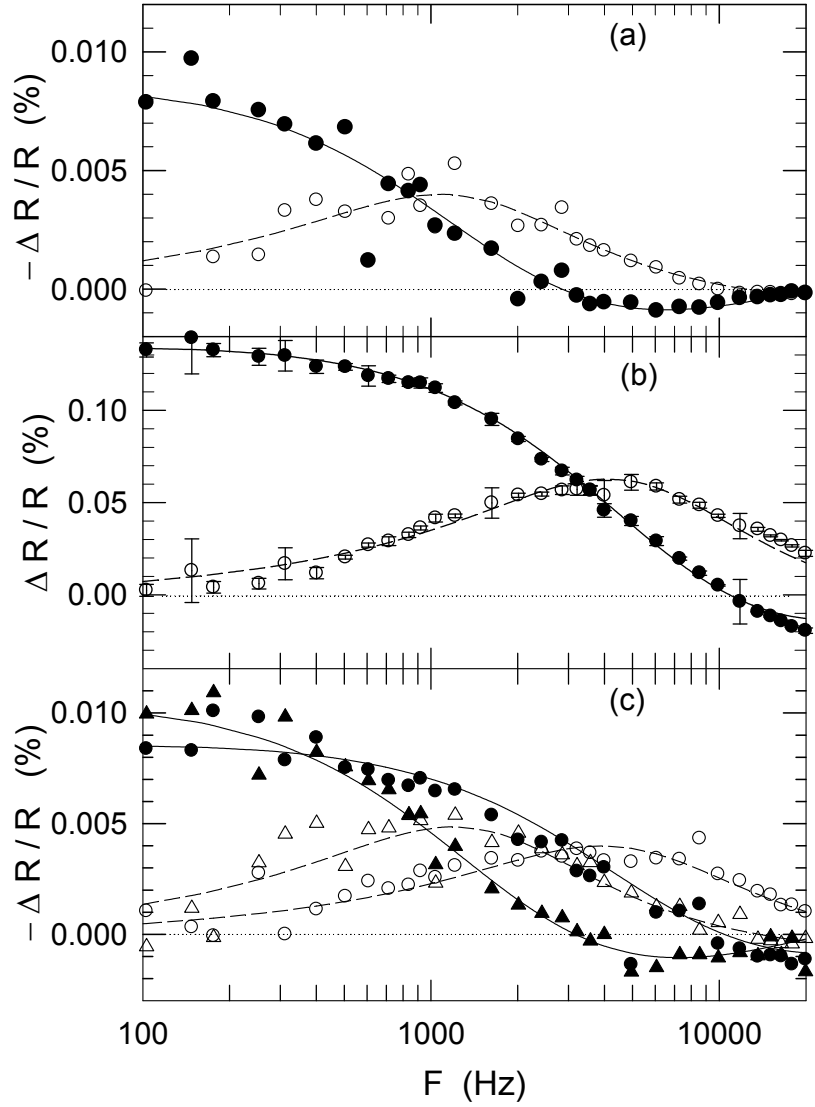

**Figure 3.** Frequency dependence of the electro-reflectance at the center of the channel of TIPS-Pn field-effect transistors, with $V_{gs0}$= -50V, and their fits to Eqtn. (4). Closed symbols and solid curves: in-phase response; open symbols and dashed curves: quadrature response. **(a)**: Transistor of Reference [7] (for which the scatter in the data shows the uncertainties). **(b)**: "Hot-spot" on another transistor, as described in the text. **(c):** Two crystals spanning the same FET channel. The circles are for the same crystal as in (b); the triangles are for a second crystal.